\documentclass[12pt]{article}
\usepackage[xdvi]{graphicx} 
\usepackage{epsfig}
\usepackage{amsfonts,amssymb,amsmath}
\usepackage{subfigure}

\newcommand{\al}{\ensuremath{\alpha}}
\newcommand{\ga}{\ensuremath{\gamma}}

\newcommand{\la}{\ensuremath{\lambda}}
\newcommand{\La}{\ensuremath{\Lambda}}

\newcommand{\p}{\ensuremath{\phi}}

\renewcommand{\d}{\ensuremath{{\rm d}}}
\newcommand{\del}{\ensuremath{\partial}}

\newcommand{\half}{\frac{1}{2}}
\newcommand{\third}{\frac{1}{3}}
\newcommand{\quarter}{\frac{1}{4}}

\newcommand{\be}{\begin{equation}}
\newcommand{\ee}{\end{equation}}
\newcommand{\m}{\mathcal}
\newcommand{\ba}{\begin{eqnarray}}
\newcommand{\ea}{\end{eqnarray}}

\begin{document}

\rightline{hep-th/0410033}
\rightline{OUTP-04/18P}
\vskip 1cm 

\begin{center}
{\Large \bf Infra-red modification of gravity from asymmetric branes}
\end{center}
\vskip 1cm 
\renewcommand{\thefootnote}{\fnsymbol{footnote}}

\centerline{\bf Antonio Padilla\footnote{a.padilla1@physics.ox.ac.uk}}
\vskip .5cm

\centerline{\it Theoretical Physics, Department of Physics}
\centerline{\it University of Oxford, 1 Keble Road, Oxford, OX1 3NP,  UK}

\setcounter{footnote}{0} \renewcommand{\thefootnote}{\arabic{footnote}}
 

\begin{abstract}
We consider a single Minkowski brane sandwiched in between two
copies of anti-de Sitter space. We allow the bulk Planck mass and
cosmological constant to differ on either side of the
brane. Linearised perturbations about this background reveal that
gravity can be modified in the infra-red. At intermediate scales, the
braneworld propagator mimics four-dimensional GR in that it has the
correct momentum dependance. However it has the wrong tensor
structure. Beyond a source dependant scale, we show that quadratic
brane bending contributions become important, and conspire to correct
the tensor structure of the propagator. We argue that even higher
order terms can consistently be ignored up to very high energies, and
suggest that there is no problem with strong
coupling. We also consider scalar and vector perturbations in the
bulk,  checking for  scalar ghosts.
\end{abstract}

\newpage
\section{Introduction}
How much do we believe in Einstein's General Theory of Relativity? We certainly don't believe it at energies beyond the Planck scale, where a quantum theory is required. In fact, experimental tests of gravity have only ever been conducted at much lower scales. For example, Newton's law has been tested down to about $0.2$ mm, or equivalently up to energies of around $10^{-3}$ eV. Solar system tests of GR are only valid outside of the Schwarzschild radius of the Sun.

Ultra-violet modifications of gravity beyond $10^{-3}$ eV occur in all
higher dimensional theories, such as String Theory. Recently, the idea
that gravity is also modified in the infra-red has received much
interest~\cite{Dvali:DGP, Kogan:multi5, Papazoglou:thesis,
Gregory:GRS,Arkani-Hamed:ghost1,Arkani-Hamed:ghost2, Padilla:bigravity}. This is entirely plausible. Again, Newton's law has only been tested up to distances of around  $10^{26}$ cm, and solar system tests of GR do not extend much beyond the size of the solar system itself.    

Infra-red modifications of gravity have some attractive features. They
can be used to tackle the cosmological constant
problem~\cite{Dvali:diluting, Barvinsky:cc} by weakening gravity at low energies. To understand this in a simple way, suppose we have Newton's constant $G$ at intermediate scales, and $G_\textrm{IR} \ll G$ in the infra-red. Even for a large cosmological constant $\La$, the contribution to the Hubble rate is considerably suppressed at large values of the scale factor
\be
H^2 \sim G_\textrm{IR}\La \ll G\La
\ee 
However, that is not to say that these models prohibit late  time inflation. On the contrary, these models generically give rise to Friedmann equations that are non-linear in $H^2$. These permit self accelerating solutions that can be tuned to the current Hubble rate~\cite{Deffayet:cosmo,Lue:accn,Dvali:power,Damour:accn, Holdom:accn,Sahni:bwde}. 

Unfortnately, these models are not without their problems. The most common of these is the vDVZ discontinuity that appears in the graviton propagator at all scales~\cite{vanDam:VDVZ, Zakharov:VDVZ}. This represents a deviation from GR that will give the wrong prediction for light bending around the Sun. In some cases, it has been argued that this discontinuity reflects a breakdown in the linearised analysis~\cite{Vainshtein:nonVDVZ,Deffayet:Vainshtein}, as opposed to a genuine sickness in the theory. Some theories predict ghost states, and therefore violate unitarity~\cite{Pilo:ghost}. A sickness as bad as this usually leads to the death of a theory. We should emphasise, however, that some models are indeed ghost-free~\cite{Dvali:DGP, Padilla:bigravity}. 

In~\cite{Padilla:asymm} we considered {\it asymmetric} branes. By
asymmetric, we mean that the gravitational parameters of the theory
can differ on either side of the brane.  There are at least two ways in
which this asymmetry might arise. Firstly, suppose we have some sort of wine bottle shaped
compactification down to 5 dimensions. In the effective theory, the
Planck scale at the fat end  of the bottle will be less than that at
the thin end. Secondly, in~\cite{Padilla:nested, Padilla:instantons}
we showed how to construct a domain wall living entirely on the
brane. In some cases~\cite{Padilla:instantons}, the Planck scale on
the brane differed on either side of the domain wall. Asymmetric
branes admit self
accelerating solutions~\cite{Padilla:asymm}. It is natural to ask,
therefore, if asymmetric brane models exhibit infra-red modifications
of gravity. 

In this paper, we will consider a positive tension brane. To the left
of the brane, the bulk is described by the Einstein-Hilbert action
with Planck mass, $M_L$, and negative cosmological constant,
$\La_L=-6/l_L^2$. To the right, we have Planck mass, $M_R$, and cosmological
constant, $\La_R=-6/l_R^2$. In general, we will take $M_L \gg M_R$ and $l_L \ll
l_R$. 

We will be particularly
interested in Minkowski branes. Because of the asymmetry, we are
allowed three classes of solution. We have (i) a Randall-Sundrum like
solution where the warp factor in the metric decays away from the
brane on both sides~\cite{Randall:RS2}, (ii) an ``inverse'' RS solution where the warp factor grows
on both sides, and (iii) a mixed solution where the warp factor decays
on one side and grows on the other.

If we place a small matter source on the brane, we can investigate
linearised perturbations about the background solutions. For the RS
solution and the mixed solution we see the introduction of a new
length scale
\be
r=\frac{M_L^3 l_L}{M_R^3}
\ee
At intermediate energies ($1/r \ll p \ll 1/l_L$), the braneworld propagator is proportional to $1/p^2$, and will reproduce Newton's
law. At lower energies ($1/l_R \ll p \ll 1/r$), the momentum
dependance changes, so that Newton's law is modified in the
infra-red. However, as in the DGP model~\cite{Dvali:DGP}, the
propagator suffers from a vDVZ discontinuity at intermediate scales,
and will predict the wrong results for light bending around the Sun.

It turns out that the vDVZ discontinuity can be removed by including
brane bending terms to {\it quadratic} order. For a source of mass
$m$, we discover yet another scale
\be
p_*=\left(\frac{M_L^3 l_L}{mr^2}\right)^\third.
\ee
For $p_* \ll p \ll 1/l_L$, the quadratic terms dominate over the
linear ones. The result is that the braneworld propagator agrees exactly with
four-dimensional General Relativity. We might be worried about the
break down of the linearised analysis at this scale. In the DGP model,
this breakdown has been linked to a strong coupling
problem~\cite{Luty:strong,Rubakov:strong, Arkani-Hamed:massive,Porrati:massive}. However,
we will argue that there is nothing to worry about. This is because
even though quadratic terms become important when  $p \sim p_*$, even higher
order terms are only important at much higher energies. It seems that
we can indeed consistently modify four-dimensional GR in the
infra-red.

The rest of this paper is organised as follows. In
section~\ref{setup}, we describe our set-up in detail, before deriving
background solutions in section~\ref{background}. In
section~\ref{sec:linear} we derive the linearised equations of motion
and solve them. We analyse the solutions and establish the existance
of the long distance scale $r$, aswell as the vDVZ discontinuity. In section~\ref{sec:radion}, we
consider scalar and vector perturbations in the bulk. Vector
perturbations only exist for the inverse RS solution. Scalar
perturbations exist whenever we include the AdS boundary. We derive the
effective action to check that these scalar fields have a well behaved
kinetic term. In other words, they are {\it not} ghosts. In
section~\ref{sec:quadratic}, we go to quadratic order in the brane
bending. We discover the new scale $p_*$, and are able to correct the
tensor structure of the propagator. In section~\ref{sec:cosmo}, we
briefly show how the mixed case admits self accelerating
solutions. Finally, section~\ref{sec:discuss} contains some
discussion. In particular, we argue that the quadratic analysis is
valid up to very high scales, and that there is no strong coupling
problem.     
\section{The set-up} \label{setup}
Consider two 5 dimensional spacetimes, $\mathcal{M}_L$ and
$\mathcal{M}_R$, separated by a domain wall. The domain wall is
a 3-brane corresponding to our universe. Our set-up is described by
the following action:
\be \label{action}
S=S_\textrm{bulk}+S_\textrm{brane}
\ee
where the contribution from the bulk is given by
\be
S_\textrm{bulk} = \sum_{i=L, R} M_i^3\int_{\mathcal{M}_i} \d^5 x
  \sqrt{-g} ~\left(R - 2\La_i\right) 
  + 2M_i^3\int_{brane} \d^4 x
  \sqrt{-\ga}~ K^{(i)}
\ee
Here, $M_i$ is the 5-dimensional Planck mass in  $\mathcal{M}_i$. We
have also included a negative cosmological constant,
$\Lambda_i=-6/l_i^2$, We will not
assume that there is $\mathbb{Z}_2$ symmetry across the brane, so that
the $M_i$ and $l_i$ can differ on either side of
the brane. The bulk metric is given by $g_{ab}$, with corresponding
Ricci scalar, $R$. $\ga_{ab}$ is the induced metric on the
brane. The extrinsic curvature of the brane in
$\mathcal{M}_i$ is given by
\be
K^{(i)}_{ab}=\ga_a^c \ga_b^d \nabla_{(c} n_{d)}
\ee 
where  $n^a$ is the unit normal to the brane in
$\mathcal{M}_i$, pointing {\it out }of $\mathcal{M}_i$. For the most
part we will not bother with the index $i$ when referring to bulk
quantities, although the reader should understand that they are
there. 

The brane part of the action is given by
\be
S_\textrm{brane} = \int_{\textrm{brane}} d^4 x
  \left(-\sigma\sqrt{-\ga} +\mathcal{L}_m\right)
\ee
where $\sigma$ is the brane tension and $\mathcal{L}_m$ describes any
additional matter.

The bulk equations of motion are given by the
Einstein equations
\be \label{Einstein}
R_{ab}-\half R g_{ab}=-\La g_{ab}
\ee
The boundary conditions at the brane are governed by the Israel
equations~\cite{Israel:eqns}. This comes from varying the action
(\ref{action}) with respect to the brane metric. Given a quantity
$Z_i$ defined in $\mathcal{M}_i$ we shall
henceforth write $\langle Z \rangle=(Z_L+Z_R)/2$, for the average across the brane, and $\Delta
Z= Z_L-Z_R$, for the difference. The brane equations of motion are
given by
\be \label{Israel}
2 \langle M^3 K_{ab} \rangle -\frac{\sigma}{6} \ga_{ab}=\half \left( T_{ab}-\third T \ga_{ab} \right)
\ee
where 
\be
T_{ab}=-\frac{~2}{\sqrt{-\ga}}\frac{\del \m{L}_m}{\del \ga^{ab}}
\ee
is the energy momentum tensor for the additional matter on the brane.

\section{Background solutions} \label{background}
In this section we will derive the metric, $\bar g_{ab}$, for the
background spacetime. These correspond to solutions to the equations
of motion when no additional matter is present. Let us introduce
coordinates $x^a=(x^\mu, z)$, and assume that the brane is located at
$z=0$. The left hand side, $\m{M}_L$ corresponds to $z<0$, whereas the
right hand side,  $\m{M}_R$ corresponds to $z>0$. In order to trust our classical analysis in the bulk, we need
to assume
\be
M > 1/l
\ee
Now seek solutions of the form
\be
\d s^2=\bar g_{ab} \d x^a \d x^b=a^2(z)\eta_{\mu\nu} \d x^\mu \d
x^\nu+\d z^2
\ee
where $\eta_{\mu\nu}$ is four-dimensional Minkowski space. The Einstein equations yield the following
\be
\left(\frac{a'}{a} \right)^2=\frac{1}{l^2}, \qquad
\frac{a''}{a}=\frac{1}{l^2}
\ee
Without loss of generality we impose the condition $a(0)=1$ so that 
\be \label{a}
a(z)= \exp\left(\frac{\theta z}{l}\right) 
\ee
where $\theta_i=\pm 1$

Given the solution (\ref{a}), the Israel equations (\ref{Israel}) impose the following condition on
the brane tension
\be \label{tension}
\Delta \left(\frac{M^3\theta}{l} \right)=\frac{\sigma}{6}
\ee
Note that this gives the usual fine-tuning of the brane tension for the Randall-Sundrum
model~\cite{Randall:RS2}, where $\Delta M=\Delta l=0$, and
$\theta_L=1=-\theta_R$.
\section{Linearised perturbations}\label{sec:linear}
We shall now consider metric perturbations about the background
solutions we have just derived. We will allow additional matter,
$T_{ab}$, to be present on the brane, but not in the bulk. We begin by
deriving the bulk equations of motion, and the boundary conditions at
the brane.
\subsection{Bulk equations of motion and boundary conditions} 
Let us
define $g_{ab}=\bar g_{ab} +\delta g_{ab}$ to be the perturbed
metric. We will work in Gaussian normal (GN) coordinates, so that
\be
\delta g_{\mu z}=\delta g_{zz}=0
\ee
Since we have no additional bulk  matter, we can take the metric
to be transverse-tracefree {\it in the bulk}. In other words,  $\delta
g_{\mu \nu}=\chi_{\mu \nu}(x, z)$, where 
\be
\del_\nu \chi^{\nu}_{\mu}=0=\chi^\mu_{\mu}
\ee
Here indices are raised and lowered with $\bar g_{\mu\nu}$. In this
choice of gauge, the linearised bulk equations of motion are given by
\be \label{bulk}
\left[ \frac{\del^2}{a^2}+\frac{\del^2}{\del z^2}-\frac{4}{l^2}
\right] \chi_{\mu\nu}(x, z)=0
\ee
Unfortunately, we no longer expect the brane to be at $z=0$. The
presence of matter causes the brane to
bend~\cite{Garriga:gravity}. In fact, we would even expect the brane
to be bent by different amounts when viewed on different sides of the
brane. In other words, the brane position is given by $z=f_i(x)$ when
viewed in $\m{M}_i$. This makes it very difficult to apply the Israel
equations (\ref{Israel}). To get round this, we need to apply a gauge
transformation in $\m{M}_L$ and another in $\m{M}_R$. The new
coordinates should also be GN, with the brane positioned at $z=0$.  

To this end, we make the following coordinate transformations
\be \label{lintran}
z \to z-f(x), \qquad x^{\mu} \to x^{\mu}-Q^{\mu}(x)+\frac{\theta l}{2}(1-a^{-2})
\del^\mu f
\ee
Note that in addition to $\theta$ and $l$, $f(x)$ and $Q^\mu(x)$ are
understood to have an invisible index $i=L, R$. In these new coordinates,
\be
\delta g_{\mu\nu}=\chi_{\mu\nu}(x, z)+2a^2 Q_{(\mu,\nu)}+\theta l \left( 1-a^2\right)\del_\mu \del_\nu f+\frac{2\theta}{l}f
\bar g_{\mu\nu}
\ee
We can evaluate this at $z=0$ to derive the brane metric
\be
\delta \ga_{\mu\nu}=\chi_{\mu\nu}(x, 0)+2Q_{(\mu, \nu)}+\frac{2\theta}{l}f
\eta_{\mu\nu}
\ee
For this to be well defined, we require that $\Delta \left(\delta
\ga_{\mu\nu}\right)=0$. Since the geometry is independant of the pure
gauge term, $2Q_{(\mu, \nu)}$, we should also demand that the remaining
part of $\delta \ga_{\mu\nu}$ is well defined. Making use of the fact
that $\chi_\mu^{\mu}=0$, this implies the
following 
\be
\Delta \chi_{\mu\nu}(x, 0)=0, \qquad \Delta Q_{(\mu, \nu)}=0, \qquad \Delta
\left(\frac{\theta f}{l} \right)=0
\ee
The last condition suggests that we introduce the function $F=\theta
f/l$. The bulk metric is now given by
\be
\delta g_{\mu\nu}=h_{\mu\nu}=\chi_{\mu\nu}(x, z)+2a^2 
Q_{(\mu, \nu)}+l^2 \left( 1-a^2\right)\del_\mu \del_\nu F+2F
\bar g_{\mu\nu}
\ee
and the brane metric by
\be
\delta
\ga_{\mu\nu}=h^\textrm{br}_{\mu\nu}=\chi_{\mu\nu}^\textrm{br}+2 Q_{(\mu, \nu)}+2F
\eta_{\mu\nu}
\ee
where $\chi_{\mu\nu}^\textrm{br}=\chi_{\mu\nu}(x, 0)$. Making use of equation (\ref{tension}), the Israel equations
(\ref{Israel}) imply the following boundary condition
\be \label{bc1}
\Delta \left[M^3 \chi'_{\mu\nu}(x, 0) \right]-\frac{\sigma}{3}
\chi^\textrm{br}_{\mu\nu}=\Sigma_{\mu\nu}(x)
\ee
where
\be
\Sigma_{\mu\nu}(x)=T_{\mu\nu}-\third T
\eta_{\mu\nu}+2\Delta (M^3 \theta l)  \del_\mu
\del_\nu F
\ee
Given that $\chi_{\mu\nu}$ is transverse-tracefree, we immediately see
that $\Sigma^{\mu}_\mu=0$. This gives
\be \label{bc2}
2\Delta (M^3 \theta l) \del^2 F=\frac{T}{3}
\ee
Here we see explicitly how matter on the brane causes it to bend.
\subsection{Solving the equations of motion}

\begin{figure}
\centering
\subfigure[The Randall-Sundrum case.]{
    \label{fig:warp:subfig:RS}
\begin{minipage}[b]{0.3\textwidth}
   \centering
   \includegraphics[width=6cm, height=6cm]{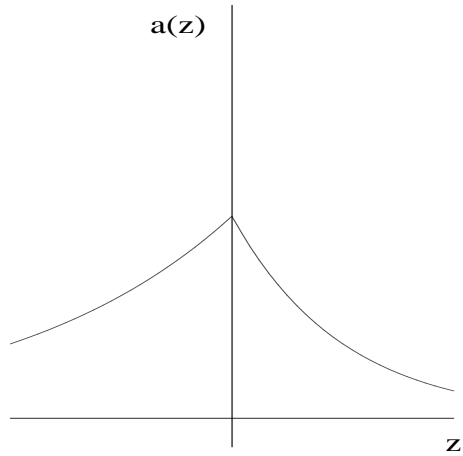}
\end{minipage}}%
\hspace{0.2\textwidth}%
\subfigure[The inverse Randall-Sundrum case.]{
   \label{fig:warp:subfig:inverse}
\begin{minipage}[b]{0.3\textwidth}
  \centering
    \includegraphics[width=6cm, height=6cm]{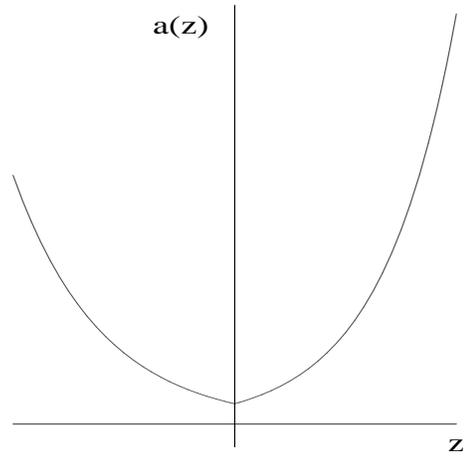} 
\end{minipage}} \\[20pt]
\subfigure[The mixed case.]{
    \label{fig:warp:subfig:mixed}
\begin{minipage}[b]{0.3\textwidth}
   \centering
     \includegraphics[width=6cm, height=6cm]{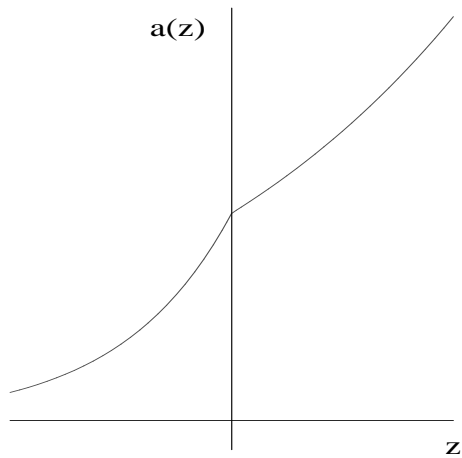}
\end{minipage}}%
\caption{The behaviour of the warp factor. For the RS
case, $a(z)$ vanishes at infinity. For the inverse RS case, $a(z)$
grows exponentially at infinity. For the mixed case, $a(z)$ vanishes at
$-\infty$ but grows exponentially at $+\infty$.} \label{fig:warp}
\end{figure}

In four dimensional momentum space, the bulk equation
(\ref{bulk}) becomes
\be 
\left[ -\frac{p^2}{a^2}+\frac{\del^2}{\del z^2}-\frac{4}{l^2}
\right] \tilde \chi_{\mu\nu}(p, z)=0
\ee
where tilde represents the Fourier transform. The general solution is 
\be
\tilde \chi_{\mu\nu}(p, z)=A_{\mu\nu}(p)I_2\left(\frac{pl}{a} \right)+B_{\mu\nu}(p)K_2\left(\frac{pl}{a} \right)
\ee
where $I_n$ and $K_n$ are modified Bessel functions of order
$n$~\cite{Abramowitz:Bessel}. Naturally, we require that
$\chi_{\mu\nu}(p, z) \to 0$ as $|z| \to \infty$. To impose this
condition we need to consider three separate cases: (i) the
{\it Randall-Sundrum case} for which $\theta_L=1=-\theta_R$,  (ii) the
{\it inverse Randall-Sundrum case} for which $\theta_R=1=-\theta_L$
and (iii) the {\it mixed case } for which $\theta_L=1=\theta_R$. In
figure~\ref{fig:warp} we can see the generic behaviour of the warp
factor in each case. 
Making use of the fact that~\cite{Abramowitz:Bessel}
\begin{eqnarray}
I_2(y) \to 0, &K_2(y) \to \infty &\textrm{as $y \to 0$} \\
I_2(y) \to \infty, &K_2(y) \to 0 &\textrm{as $y \to \infty$}
\end{eqnarray}
we see that the boundary condition at infinity is satisfied if and
only if
\be \label{solution}
\tilde \chi_{\mu\nu}(p, z)=\frac{C_2(pl/a)}{C_2(pl)} ~\tilde \chi_{\mu\nu}^\textrm{br}(p)
\ee
where
\be
C_2={\begin{cases}K_2 & \textrm{for RS case} \\
~I_2 & \textrm{for inverse RS case} \end{cases}}
\ee
and for the mixed case $C_2^L=K_2$ and $C_2^R=I_2$.

The Fourier transformed boundary conditions (\ref{bc1}) and
(\ref{bc2}) are given by
\ba
\Delta \left[M^3 \tilde \chi'_{\mu\nu}(p, 0) \right]-\frac{\sigma}{3}
\tilde \chi^\textrm{br}_{\mu\nu}&=&\tilde \Sigma_{\mu\nu}(p)
\label{FTbc1}\\
2\Delta (M^3 \theta l) p^2 \tilde
F &=&-\frac{\tilde T}{3}\label{FTbc2}
\ea
Given that~\cite{Abramowitz:Bessel} 
\be
yI_2'(y)+2I_2(y)=yI_1(y) \qquad yK_2'(y)+2K_2(y)=-yK_1(y)
\ee
we can insert the solution (\ref{solution}) into (\ref{FTbc1}) to give
\be \label{chi}
\tilde \chi_{\mu\nu}^{\textrm{br}}=\frac{\tilde \Sigma_{\mu\nu}(p)}{R(p)}
\ee
where
\be
R(p)=-p\Delta\left[M^3 \theta \frac{C_1(pl)}{C_2(pl)}
\right]
\ee
Here
\be
C_1={\begin{cases}-K_1 & \textrm{for RS case} \\
~~I_1 & \textrm{for inverse RS case} \end{cases}}
\ee
and for the mixed case $C_1^L=-K_1$ and $C_1^R=I_1$. Note that we have
also used equation (\ref{tension}) in deriving equation (\ref{chi}).

In momentum space, the metric on the brane is given by
\be
\tilde h_{\mu\nu}^\textrm{br}=\tilde \chi_{\mu\nu}^\textrm{br}+2ip_{(\mu}\tilde Q_{\nu)}+2 \tilde F \eta_{\mu\nu}.
\ee
We can choose $\tilde Q_{\mu}(p)$ to cancel off the term proportional to
$p_\mu p_\nu \tilde F$ in $\tilde \Sigma_{\mu\nu}(p)$. This leaves
\be \label{h}
\tilde h_{\mu\nu}^\textrm{br}=\frac{1}{R(p)} \left[\tilde T_{\mu\nu}-\third
\tilde T \eta_{\mu\nu}\right]-\frac{1}{6\alpha p^2} \tilde T \eta_{\mu\nu}
\ee
where
\be
\alpha=\half \Delta (M^3 \theta l)
\ee
This all makes sense as  long as $R(p)$ and $\alpha$ do not vanish in
$p>0$.  

At some energy scale, the
solution (\ref{h}) should agree with the corresponding result from
4-dimensional Einstein gravity
\be \label{prop}
\tilde h_{\mu\nu}=\frac{1}{m_{pl}^2 p^2} \left[\tilde T_{\mu\nu}-\half
\tilde T \eta_{\mu\nu}\right]
\ee
This requires there to be an energy scale for which $R(p) \sim \alpha p^2$ and $\alpha=m_{pl}^2
>0$. We shall now investigate whether or not such a scale exists.

At energies that are small compared with the bulk curvature, we have
$pl \ll 1$. In this limit we can approximate the Bessel functions as
follows~\cite{Abramowitz:Bessel}  
\be
I_1(pl) \sim \half pl, \qquad I_2(pl) \sim \frac{1}{8} (pl)^2, \qquad
K_1(pl) \sim \frac{1}{pl}, \qquad K_2(pl) \sim \frac{2}{(pl)^2}
\ee
At high energies, $pl \gg 1$, we have
\be
\frac{I_1(pl)}{I_2(pl)} \sim 1, \qquad \frac{K_1(pl)}{K_2(pl)} \sim 1 
\ee
For the RS case ($\theta_L=1=-\theta_R$) we see that
\be
R(p)=2p\Big \langle M^3 \frac{K_1(pl)}{K_2(pl)}~ \Big
\rangle \sim {\begin{cases}\alpha p^2 & \textrm{for $p \ll 1/l_R$} \\
\half M_L^3 l_L p^2+M_R^3 p  & \textrm{for $1/l_R \ll p \ll
1/l_L$} \\
2\langle M^3 \rangle p  & \textrm{for $p \gg 1/l_L $} \end{cases}}
\ee
where $\alpha=\langle M^3 l \rangle$. Note that we have assumed $l_R
\gg l_L$.  

At first glance, its appears that we only have Einstein gravity when $p \ll 1/l_R$.  However, if $M_L^3l_L \gg
M_R^3l_R$, $R(p)
\sim \half M_L^3l_L p^2$ for $1/l_R \ll p \ll
1/l_L$. Furthermore, we now have $\alpha \approx  \half
M_L^3l_L$. This means that we also have Einstein gravity whenever
$1/l_R \ll p \ll
1/l_L$. In any case, there is no modified gravity in the infra-red.

For the inverse RS case ($\theta_R=1=-\theta_L$), we see that
\be
R(p)=2p\Big \langle M^3 \frac{I_1(pl)}{I_2(pl)}~ \Big
\rangle \sim {\begin{cases}8 \langle M^3/l \rangle & \textrm{for $p \ll 1/l_R$} \\
M_L^3 p+4\frac{M_R^3}{l_R}   & \textrm{for $1/l_R \ll p \ll
1/l_L$} \\
2\langle M^3 \rangle p  & \textrm{for $p \gg 1/l_L $} \end{cases}}
\ee
and $\alpha=- \langle M^3l \rangle$. Einstein gravity cannot be
produced at any scale.

Finally, for the mixed case ($\theta_L=1=\theta_R$), we see that 
\be
R(p) =p\left[
M_L^3\frac{K_1(pl_L)}{K_2(pl_L)}+M_R^3\frac{I_1(pl_R)}{I_2(pl_R)}
\right] \sim {\begin{cases}\half M_L^3l_Lp^2+4\frac{M_R^3}{l_R} & \textrm{for $p \ll 1/l_R$} \\
\half M_L^3l_Lp^2 +M_R^3p& \textrm{for $1/l_R \ll p \ll
1/l_L$} \\
2\langle M^3 \rangle p  & \textrm{for $p \gg 1/l_L $} \end{cases}}  \\
\ee
and $\alpha = \half \Delta(M^3l)$. Again, at first glance  its appears that
Einstein gravity can never be reproduced. However, as in the RS case, if $M_L^3l_L \gg
M_R^3l_R$, we have Einstein gravity for $1/l_R \ll p \ll
1/l_L$. The difference here is that gravity {\it can} be modified in
 the far infra-red region, $p \ll 1/l_L$. Unfortunately, as we shall
see in section \ref{sec:radion}, the mixed case contains a
radion ghost whenever $\alpha >0$. Here $\alpha \approx \half M_L^3
l_L >0$.

These results are a little disappointing. We can only get Einstein
gravity in the RS case, without any modifications in the
infra-red. This is nothing new~\cite{Randall:RS2}. 

However, all is not lost! Consider the RS case and the mixed case for
$p \gg 1/l_R$. Now suppose  $M_L^3l_L \ll
M_R^3l_R$, in contrast to what we discussed earlier. This gives
\be
\alpha \approx {\begin{cases} ~~\half M_R^3 l_R &\textrm{for the RS
case} \\
-\half M_R^3 l_R &\textrm{for the mixed
case} \end{cases}}
\ee
Since $\alpha<0$ for the mixed case, there is no radion ghost. In both
cases, we see the introduction of a new length scale
\be
r=\frac{M_L^3 l_L}{M_R^3} \ll l_R
\ee
We now have
\be
R(p) \sim {\begin{cases}
M_R^3 p & \textrm{for $1/l_R \ll
p \ll 1/r$} \\
 \half M_L^3 l_L p^2 & \textrm{for $1/r \ll
p \ll 1/l_L$} \\
M_L^3 p & \textrm{for $p \gg 1/l_L$}
\end{cases}}
\ee
Note that we have also assumed that $M_L \gg M_R$. In these limits, the brane bending term, $\tilde T/\alpha p^2$ is always much less than the remaining terms in $\tilde h_{\mu\nu}^\textrm{br}$. This implies that
\be \label{sol}
\tilde h_{\mu\nu}^\textrm{br} \approx \frac{1}{R(p)} \left[\tilde T_{\mu\nu}-\third
\tilde T \eta_{\mu\nu}\right]
\ee
In the intermediate regime ( $1/r \ll
p \ll 1/l_L$), our solution
has the correct momentum dependence, but the wrong tensor
structure. As we move into the infra-red ($1/l_R \ll
p \ll 1/r$), or the ultra-violet ($p \gg 1/l_L$) our momentum dependence
changes so that gravity is modified. 

Of course, our solution (\ref{sol}) no longer makes sense when
$\tilde h_{\mu\nu}^\textrm{br}$ is of order one. For a source of mass, $m$,
note that $|\tilde T| \sim mp^3$. This means that $\tilde
h_{\mu\nu}^\textrm{br}$ schematically goes like $mp^3/R(p)$. This
becomes of order one when $p \sim p_\textrm{cut-off}$. For very large masses
($m>M_L^3l_L r$), this cut-off occurs in the infra-red. Otherwise we have
\be \label{cut-off}
 p_\textrm{cut-off}={\begin{cases} {M_L^3l_L}/{m} & \textrm{for
$M_L^3 l_L^2<m <M_L^3l_L r$} \\
\left({M_L^3}/{m} \right)^{{1}/{2}}  &\textrm{for $m<M_L^3
l_L^2$} \end{cases}} .
\ee
Note that the upper value in (\ref{cut-off}) lies in the intermediate
energy range, whereas the lower value lies in the ultra-violet.

We might think that this cut-off represents the scale at which the
linearised theory breaks down. Certainly, the linearised theory makes
no sense beyond the cut-off, but is it really valid beforehand? In the
DGP model~\cite{Dvali:DGP}, the linearised theory actually breaks down
much sooner than expected. This is because higher order terms become
important~\cite{Deffayet:Vainshtein}. Curiously, these non-linear
terms can sometimes correct   the problems with the tensor structure at
intermediate energies~\cite{Lue:strings, Gruzinov:mass, Lue:leakage, Tanaka:DGP, Middleton:DGP}. In
section~\ref{sec:quadratic}, we will see that the same thing happens here.

\section{Scalar and vector perturbations} \label{sec:radion}
We shall now discuss scalar and vector perturbations in the bulk, when
there is no additional matter on the brane ($T_{ab}=0$).
\subsection{Scalar perturbations and ghosts}
Braneworld models that exhibit infra-red modifications of gravity
often contain ghosts (see,  for example~\cite{Gregory:GRS,
Pilo:ghost}). Typically, these ghosts appear in the scalar sector. We will be forced to eliminate any model that contains a
ghost-like scalar field in the effective theory.

Consider the following scalar perturbations about the background,
$\bar g_{ab}$, 
\be
\delta g_{\mu\nu}=a^2\left( 2\del_\mu \del_\nu E+2A\eta_{\mu\nu}
\right), \qquad \delta g_{\mu z}=\del_\mu B, \qquad \delta g_{zz}=2\p
\ee
Under the scalar gauge transformations
\be
x^\mu \to x^\mu +\del^\mu \xi, \qquad z \to z+ \xi_z
\ee
the scalar fields transform as follows
\be
E \to E-\xi,\qquad  A \to A-\frac{\theta}{l} \xi_z, \qquad B \to
B-(\xi_z+a^2 \xi'), \qquad \p \to \p-\xi_z'
\ee
It is easy to check that the following quantities are gauge invariant
\be
X=\phi-l\theta A', \qquad Y=B-a^2 E'-l\theta A
\ee
The boundary condition at infinity requires there to be no physical
perturbation there. This amounts to the gauge invariants $X$ and $Y$
vanishing at infinity. We can see this most easily by choosing the
gauge $E=A=0$, so that $X=\p$ and $Y=B$.

In terms of the gauge invariants, the bulk equations of motion (\ref{Einstein}) are given by
\ba
0&=&\del_\mu \del_\nu \left[ Y'+\frac{2 \theta}{l}Y-2X
\right]+\frac{\theta}{l}a^2 \eta_{\mu\nu}\left[\frac{\del^2Y}{a^2}+X'+\frac{8\theta}{l}X\right] \\
0 &=& \frac{3\theta}{l}\del_\mu X \\
0&=& \frac{\del^2Y'}{a^2}-  \frac{\del^2X}{a^2}+\frac{4\theta}{l}X'+\frac{8}{l^2}X
\ea
The solution is
\be
X=0, \qquad Y=\frac{U(x)}{a^2}
\ee
where $\del^2U=0$. In principle, $U$ can differ on
either side of the brane. 

Recall that we require $X, Y \to 0$ as $|z| \to \infty$. For the RS
case, this requires that $U_L(x)=U_R(x)=0$. For the inverse RS
case, $U(x)$ can be non-zero on both sides of the brane. Finally, for
the mixed case, we must have $U_L(x)=0$, although $U_R(x)$ can be
non-zero.

In order to apply the boundary conditions near the brane we have to
choose a gauge. We can choose GN gauge whilst keeping the brane
position fixed at $z=0$. 
We now have $B=\p=0$, and 
\be
E=\quarter \theta l U(x) (a^{-4}-1)+\half l^2 V(x)
(a^{-2}-1)+W(x), \qquad A=V(x)
\ee
We can evaluate our solution at $z=0$ to derive the brane metric
\be
\delta \gamma_{\mu \nu}=2\del_\mu\del_\nu W+2V\eta_{\mu\nu}
\ee
The pure gauge part, $2\del_\mu\del_\nu W$, and the remainder,
$2V\eta_{\mu\nu}$, should both be well defined. This means that $\Delta
V=\Delta W=0$. In particular, since $\Delta W=0$, it is easy to see
that $W$ can be continuously gauged away on both sides of the brane. We
therefore set $W=0$.

The Israel equations (\ref{Israel}) are now given by
\be
2\del_\mu \del_\nu \Delta (M^3 E')=0
\ee
where we have made use of equation (\ref{tension}). This implies that
\be \label{BC1}
\Delta(M^3U)=-\Delta(M^3\theta l) V=-2\alpha V
\ee
For the RS case ($\theta_L=1=-\theta_R$), we have deduced that
$U_L=U_R=0$. Since $\alpha=\langle M^3 l \rangle \neq 0$, equation
(\ref{BC1}) clearly implies that $V=0$. This means that there is no scalar
perturbation in the RS case.

In contrast, for the mixed case and the inverse RS case, we conclude
that there is at least one remaining degree of freedom, $V$, say. This
means that there is at least one scalar perturbation, which we will
refer to as the radion.

We need to check whether or not the radion is a ghost. This involves
calculating the radion effective action. Again, we need to choose a
gauge. Our gauge choice must correspond to the brane being fixed at
$z=0$, and the perturbation vanishing at infinity.

Let us begin by choosing the gauge $E=A=0$, for which
\be
B=\frac{U}{a^2}, \qquad \p=0
\ee
This vanishes at infinity. Near the brane, we begin in GN gauge, but
transform to $E=A=0$ by choosing
\be
\xi= \quarter \theta l U(x) (a^{-4}-1)+\half l^2 V(x)
(a^{-2}-1), \qquad \xi_z=l\theta V
\ee
The brane is now positioned at $z=l\theta V$. We need to move it back
to $z=0$ without introducing a perturbation at infinity. This can be
done by choosing $\xi_z=P(z)$, where $P(0)=-l\theta V$ and $P \to 0$
as $|z| \to \infty$. The bulk perturbation is now given by the
following
\be
E=0, \qquad A=-\frac{\theta}{l}P(z), \qquad B=\frac{U}{a^2}
-P(z), \qquad \p=-P'(z)
\ee
To quadratic order, the effective action is
\be
S_\textrm{eff}=-\half \int_\textrm{bulk} \sqrt{-g}M^3 h^{ab} \delta
E_{ab}-\half \int_\textrm{brane} \sqrt{-\ga} h^{ab}\delta
\Theta_{ab}
\ee
where $\delta
E_{ab}$ and $\delta
\Theta_{ab}$ are the expansions, to linear order, of the bulk and
boundary equations of motion respectively
\ba
E_{ab}&=&R_{ab}-\half Rg_{ab}+\La g_{ab} \\
\Theta_{ab}&=& 2 \langle M^3 (K_{ab}-K \ga_{ab}) \rangle
+\frac{\sigma}{2} \ga_{ab}
\ea
We find that
\ba
S_\textrm{eff}&=&-\frac{3}{2}\int d^4x \left
[ \frac{M^3_L\theta_L}{l_L}\del^2 U_L\int_{-\infty}^{0^-} dz
P'(z)+\frac{M^3_R\theta_R}{l_R}\del^2 U_R\int^{\infty}_{0^+} dz
P'(z) \right] \nonumber \\
&=&\frac{3}{2}\int d^4x V\del^2 \Delta(M^3 U)=-3\alpha \int d^4x V
\del^2 V
\ea
where we have made use of equation (\ref{BC1}). We immediately see that
the radion is a ghost whenever $\alpha>0$. This is not a problem for
the inverse RS case, as $\alpha=-\langle M^3 l\rangle <0$. However,
as we saw in the last section, this {\it can} be a problem for the
mixed case.

\subsection{Vector perturbations}
Now consider vector perturbations in the bulk. Since brane bending
represents a scalar fluctuation we can set the brane to be at
$z=0$. In general, the vector perturbation is given by
\be
\delta g_{\mu\nu}=2a^2 F_{(\mu, \nu)}, \qquad \delta g_{\mu z}=B_\mu,
\qquad \delta g_{zz}=0
\ee
where $F_\mu$ and $B_\mu$ are divergence-free.

Under the vector gauge transformation, 
\be
x^\mu \to x^\mu +\xi^\mu, \qquad z\to z
\ee
the vector fields transform as follows
\be
F_\mu \to F_\mu -\xi_\mu, \qquad B_\mu \to B_\mu-a^2 \xi_\mu'
\ee
where $\xi_\mu=\eta_{\mu\nu} \xi^\nu$. Note that $\xi^\mu$ is
divergence-free. It is clear that we have the following gauge
invariant
\be
C_\mu=B_\mu-a^2 F_\mu'
\ee
The bulk equations of motion (\ref{Einstein}) are given by
\ba
0 &=& \left( \del_z+\frac{2\theta}{l} \right)C_{(\mu, \nu)} \\
0 &=& -\half \frac{\del^2}{a^2} C_\mu
\ea
These have the following solution 
\be
C_{\mu}=\frac{\lambda_\mu (x)}{a^2}
\ee
where $\del^2 \la_\mu=0$. Of course $\la_{\mu}$ can differ on either side of the brane. In analogy
with the scalar gauge invariant, we require that $C_\mu \to 0$ as $|z|
\to \infty$. 
 For the RS
case, this requires that $\la_{\mu}^L=\la_{\mu}^R=0$. For the inverse RS
case, $\la_{\mu}$ can be non-zero on both sides of the brane. Finally, for
the mixed case, we must have $\la_{\mu}^L=0$, although
$\la_{\mu}^R$ can, in principle, be
non-zero.

To apply the boundary conditions we go to GN gauge ($B_\mu=0$), so that
\be
F_{\mu}=\quarter\theta l \la_{\mu}( a^{-4}-1 ) +G_{\mu}(x)
\ee
Continuity at the brane implies that $\Delta G_{\mu}=0$, so we can
continuously gauge away $G_\mu$. The Israel equations (\ref{Israel})
give
\be
2 \Delta \left[ M^3 F_{(\mu, \nu)}' \right]=0
\ee
This implies that
\be
\Delta \left[M^3 \la_{\mu} \right]=0
\ee
It is easy to see that if $\la_{\mu}$ is zero {\it any}where, it is
zero {\it every}where. We conclude that  vector perturbations can only exist for the inverse RS case.
\section{Beyond the linearised analysis} \label{sec:quadratic}
 At the end of section \ref{sec:linear}, we entertained the
 possibility of the linearised analysis breaking down at an
 unexpectedly low scale. In this section, we will show that this is indeed
 the case. We will go to quadratic order to gain insight into the full
 non-linear theory.

Let us begin with the transverse-tracefree + GN gauge in the bulk
\be
\delta g_{\mu\nu}=\chi_{\mu\nu}, \qquad \delta g_{\mu z}=\delta g_{zz}=0
\ee
The brane is now bent. We need to fix the position of the brane to be
at $z=0$, whilst remaining in GN gauge {\it to quadratic order}. To
this end, we make the following
coordinate transformation
\be
z \to z-f(x)-\frac{\theta l}{4} (1-a^{-2}) \del_\mu f \del^\mu f 
\ee
\be
x^{\mu} \to x^{\mu}-Q^{\mu}(x)+\frac{\theta l}{2}(1-a^{-2})
\left[\del^\mu f+\del_{\nu}f \del^{\mu}Q^{\nu}+\frac{2\theta}{l}f
\del^\mu f \right]+\int_0^z \chi^{\mu\nu} \del_\nu f
\ee
Note that this agrees with the linearised transformations
(\ref{lintran}) to first order. The bulk metric is now given by
\be \label{quadh}
\delta g_{\mu\nu}=h_{\mu\nu}=\chi_{\mu\nu}(x, z)+2a^2 
Q_{(\mu, \nu)}+l^2 \left( 1-a^2\right)\del_\mu \del_\nu F+2F
\bar g_{\mu\nu}-l^2 \del_{\mu}F  \del_{\nu}F+\delta h_{\mu\nu}
\ee
where $F=\theta f/l$. The terms in $\delta h_{\mu\nu}$ are always much
smaller than the other terms in $h_{\mu\nu}$, so we will neglect
them. For the linearised analysis to be valid, we require that the
linear terms in (\ref{quadh}) are much larger than the remaining
quadratic term, $l^2 \del_{\mu}F  \del_{\nu}F$.  

Suppose $p \gg 1/l_R$, $M_L \gg M_R$ and $M_L^3
l_L \ll M_R^3l_R$. For a source of mass $m<M_L^3l_Lr$, we have shown
that the linearised analysis certainly breaks down when $p \sim
p_\textrm{cut-off}$, where $p_\textrm{cut-off}$ is given by equation
(\ref{cut-off}). We will now show that the quadratic term,  $l^2
\del_{\mu}F  \del_{\nu}F$, actually becomes important much sooner. 

From section~\ref{sec:linear}, we know that the linear terms go like  $mp^3/R(p)$.  Schematically,
the quadratic term becomes important when $p^2 l^2 F^2 \sim
mp^3/R(p)$. To identify when this happens we need to know the size of
$F$. From equation (\ref{FTbc2}), we see that
\be
|F_L|=|F_R| \sim \frac{|T|}{|\al| p^2} \sim  \frac{|T|}{M^3_R l_R p^2} \sim  \frac{mp}{M^3_R l_R}
\ee
It turns out that the quadratic term becomes important when $p \sim
p_* \ll p_\textrm{cut-off}$. For the remainder of this section we will
restrict our attention to sources with the following mass
\be
M_L^3l_L^4/r^2 <m< M_L^3 l_L r
\ee
When $r$ is large, this covers a huge range of masses. By taking our
mass to lie in this range we ensure that $1/r < p_* <1/l_L$. Specifically,
\be
p_*=\left(\frac{M_L^3l_L}{mr^2} \right)^\third
\ee
We conclude that the linearised analysis only makes sense for $p \ll p_* \ll
p_\textrm{cut-off}$.  For $p_* \ll p \ll p_\textrm{cut-off}$, we
cannot ignore the quadratic term, $l_R^2 \del_{\mu}F_R
\del_{\nu}F_R$. Let us proceed with this in mind.

It is convenient to decompose the remaining quadratic as follows
\be
-l^2 \del_{\mu}F  \del_{\nu}F=\mathcal{F}_{\mu\nu}+ 2 F_{(\mu,
\nu)}+2\del_\mu\del_\nu E+2A\eta_{\mu\nu}
\ee
where $\mathcal{F}_{\mu\nu}$ is transverse-tracefree and $F_\mu$ is
divergence free. It is easy to show that
\be
\del^2 A=\frac{~l^2}{6} \left[ \del^\mu\del^\nu -\eta^{\mu\nu}\del^2
\right]\del_\mu F \del_\nu F
\ee 
Since $\mathcal{F}_{\mu\nu}$ is transverse-tracefree, let us absorb it
into a redefinition of $\chi_{\mu\nu}$. Furthermore, let
\be
Q_{\mu} \to Q_{\mu}-F_{\mu}-\del_{\mu} E
\ee
so that the bulk metric is given by
\begin{multline}
h_{\mu\nu}=\chi_{\mu\nu}(x, z)+2a^2 
Q_{(\mu, \nu)}+l^2 \left( 1-a^2\right)\del_\mu \del_\nu F+2F
\bar g_{\mu\nu} \\+2(1-a^2)\left[F_{(\mu, \nu)}+\del_\mu\del_\nu E
\right] +2A\eta_{\mu\nu}
\end{multline}
We can evaluate this at $z=0$ to derive the brane metric,
\be
\delta \ga_{\mu\nu}=h_{\mu\nu}^\textrm{br}=\chi_{\mu\nu}(x, 0)+2Q_{(\mu, \nu)}+2(F+A)\eta_{\mu\nu}
\ee
The pure gauge part, $2Q_{(\mu, \nu)}$, and the remainder must be well
defined. Making use of the fact that $\chi^{\mu}_\mu=0$, we find
\be \label{quadbc1}
\Delta  \chi_{\mu\nu}(x, 0)=0, \qquad  \Delta Q_{(\mu, \nu)}=0, \qquad
\Delta (F+A)=0
\ee
It turns out that the Israel equations take very nearly the same form
as in the linearised theory, the only subtely being that $\Delta F\neq
0$. Once again we have equation (\ref{bc1}) but with
\be \label{quadbc2}
\Sigma_{\mu\nu}(x)=T_{\mu\nu}-\third T
\eta_{\mu\nu}+2\del_\mu\del_\nu\Delta(M^3 \theta l F)
\ee
Note that we have used the fact that
\be
F_{(\mu, \nu)}+\del_\mu\del_\nu E+A \eta_{\mu\nu} \sim p^2l^2 F^2 \ll
p^2l^2 F \sim l^2 \del_\mu\del_\nu F
\ee
Given that $\chi_{\mu\nu}$ is transverse-tracefree, we have that
\be \label{quadbc3}
2\del^2 \Delta (M^3 \theta lF)=\frac{T}{3}
\ee
In momentum space, the metric on the brane is given by
\be
\tilde h_{\mu\nu}^\textrm{br}=\tilde
\chi_{\mu\nu}^\textrm{br}+2ip_{(\mu}\tilde Q_{\nu)}+2 (\tilde F+\tilde
A )\eta_{\mu\nu}.
\ee
As in section \ref{sec:linear}, we see that
\be
\tilde \chi_{\mu\nu}^\textrm{br}=\frac{\tilde
\Sigma_{\mu\nu}(p)}{R(p)}
\ee
We choose $\tilde Q_\mu(p)$ to cancel off the term proportional to
$p_\mu p_\nu\Delta(M^3 \theta l \tilde F)$ in $\tilde
\Sigma_{\mu\nu}(p)$. This leaves
\be
\tilde h_{\mu\nu}^\textrm{br}=\frac{1}{R(p)} \left[\tilde T_{\mu\nu}-\third
\tilde T \eta_{\mu\nu}\right]+2(\tilde F+\tilde
A )\eta_{\mu\nu}
\ee
For the RS case and the mixed case, recall that
\be
R(p) \sim {\begin{cases}
M_R^3 p & \textrm{for $1/l_R \ll
p \ll 1/r$} \\
 \half M_L^3 l_L p^2 & \textrm{for $1/r \ll
p \ll p_\textrm{uv}$} \end{cases}}
\ee
provided $M_L \gg M_R$ and $M_L^3 l_L \ll M_R^3 l_R$. Here
\be
p_\textrm{uv}=\textrm{min} \{1/l_L, p_\textrm{cut-off} \}=\textrm{min}
\{1/l_L, M^3_Ll_L/m \}.
\ee 
Our goal is to
reproduce Einstein gravity at certain scales. $R(p)$ already has the correct momentum
dependance  when $1/r \ll
p \ll p_\textrm{uv}$. To ensure that our solution also has the right tensor
structure (see equation (\ref{prop})), we require that
\be \label{F+A}
\tilde F+\tilde A =-\frac{\tilde T}{6 M_L^3 l_L p^2}
\ee
We will now see that this is indeed the case when $p_* \ll p \ll p_\textrm{uv}$.

For $p \ll p_*$, we know that the linearised analysis can be trusted. However, for $p_* \ll p \ll p_\textrm{uv}$, we have
\be \label{ineq}
F_L \gg A_L, \qquad F_R \ll A_R
\ee
Making use of  boundary condition (\ref{quadbc1}), we deduce that 
\be
F_L \approx F_L+A_L=F_R+A_R \approx A_R
\ee
Now consider the trace equation (\ref{quadbc3}). In principle, we have
two possibilities: (i) $M^3_L l_L |F_L| \gg  M^3_R l_R |F_R|$ or (ii)
$M^3_L l_L |F_L| \ll  M^3_R l_R |F_R|$. For  $p_* \ll p \ll
p_\textrm{uv}$, it turns out that case (i) is
the only one to give a self-consistent solution. The
trace equation (\ref{quadbc1}) now implies that
\be
\tilde F_L \approx -\frac{\tilde T}{6 M_L^3 l_L p^2}
\ee
so that equation (\ref{F+A}) holds. 

To briefly sum up, we have shown that the RS case and the mixed case
mimic four-dimensional General Relativity when  $p_* \ll p \ll
p_\textrm{uv}$. This happens because the linearised results break down,
and we have to include quadratic contributions. At lower energies, the
linearised results hold, and we deviate from four dimensional GR. At first, the tensor structure of our propagator
changes. As we go even deeper into the infra-red, our momentum
dependance also changes.

It is important to ask whether or not the radion analysis of section
\ref{sec:radion} is significantly affected by non-linear effects. The
answer is clearly ``no''. This is because the radion field is massless
$p^2=0$, so the linearised analysis can be trusted. Recall that there
is no  radion in the RS case, and the radion is {\it not} a ghost for
the mixed case when  $M^3_L l_L< M^3_R l_R$. We conclude that the
radion is not a problem here.

\section{A word on cosmology} \label{sec:cosmo}

Infra-red modifications of gravity generically have far reaching
implications for cosmology, particularly at late times. In this section, we
will briefly examine the cosmological solutions that arise in our
models. We will omit most of the details, as they can be found
in~\cite{Padilla:asymm}. We should note that in~\cite{Padilla:asymm},
$\theta_R$ is defined with the opposite sign to here.

Let us consider a spatially flat Friedmann-Robertson-Walker brane
sandwiched in between two copies of anti-de Sitter space. The dynamics
of the brane are governed by the following equation~\cite{Padilla:asymm}
\be \label{FRW}
\Delta \left[M^3 \theta \sqrt{\frac{1}{l^2}+H^2}\right]=\frac{\rho}{6}
\ee
where $H$ is the Hubble parameter on the brane, and $\rho$ is the
energy density. Here $\rho$ is made up of brane tension and additional
matter
\be
\rho=\sigma+\rho_m
\ee
Note that $\sigma$ is given by equation (\ref{tension}). For $1/l_R \ll H \ll 1/l_L$,  equation (\ref{FRW}) approximates to
\be \label{appFRW}
\theta_L H^2-2\theta_R \frac{|H|}{r} \approx \frac{\rho_m}{3M_L^3 l_L}
\ee
At early times, $H \gg 1/r$, so that
\be
H^2 \approx  \frac{\theta_L\rho_m}{3M_L^3 l_L}
\ee
For the RS and the mixed case, $\theta_L=1$, so this simply gives the
standard 4d cosmology with $m_{pl}^2=\half M_L^3 l_L$. 

We might expect to see a de
Sitter phase of expansion at later times, as $\rho_m \to 0$. For the
RS case ($\theta_L=-\theta_R=1$), it is clear from equation (\ref{FRW}) that this will not
happen: $H^2=0$ is the only
solution for  $\rho_m = 0$. In contrast, for the mixed
case ($\theta_L=\theta_R=1$), the solution $H\approx 2/r$ is permitted.

We conclude that late time de Sitter expansion occurs for the mixed
case, but not the RS case. The RS result is a surprise as we have
modified gravity in the range $1/l_R \ll p \ll 1/r$.

\section{Discussion} \label{sec:discuss}
We have considered in detail perturbations about a flat brane embedded
in between two copies of anti-de Sitter space. We have abandoned
$\mathbb{Z}_2$ symmetry across the brane, giving rise to three
distinct cases: (i) the RS case where the warp factor decays away from
the brane on both sides, (ii) the inverse RS case where the warp factor
grows on both sides and (iii) the mixed case where the warp factor
decays to the left of the  brane, and grows to the right.
 
Let us now summarize our main results. The inverse RS case is a dead
end: four dimensional Einstein gravity can {\it never} be
achieved. In contrast, take the RS case and the mixed case when $M_L
\gg M_R$ and $M_L^3 l_L \ll M_R^3 l_R$. Consider a source of mass, $m
\in ( M_L^3 l_L^4/r^2, M_L^3 l_L r)$, where $r=M_L^3 l_L/M_R^3$. The metric on the brane is given by
\be \label{summary}
\tilde h_{\mu\nu}^\textrm{br}={\begin{cases}\frac{1}{m_{pl}^2p^2} \left[ \tilde T_{\mu\nu}-\half \tilde T \eta_{\mu\nu} \right] & \textrm{when $p_* \ll p \ll p_\textrm{uv}$}\\
\frac{1}{m_{pl}^2p^2} \left[ \tilde T_{\mu\nu}-\third \tilde T \eta_{\mu\nu} \right] & \textrm{when $1/r \ll p \ll p_*$}\\
\frac{1}{M_R^3p} ~\left[ \tilde T_{\mu\nu}-\third \tilde T \eta_{\mu\nu} \right] & \textrm{when $1/l_R \ll p \ll 1/r$} \end{cases}}
\ee
where $m_{pl}^2=\half M^3_L l_L$, $p_*=(M^3_Ll_L/mr^2)^{1/3}$, and  
$p_\textrm{uv}=\textrm{min} \{ 1/l_L, M^3_L l_L/m \}$. Gravity on the
brane mimics four-dimensional GR at intermediate scales, but is
modified in the infra-red. This modification begins with a change in
the tensor structure of the propagator. At even lower energies the
momentum dependance also changes. Crucially, these models are free
from ghosts. 

We should also note that a cosmological brane in the mixed case will
approach a de Sitter phase at late times, even when there is no
additional matter. This does not happen for the RS case.

To illustrate our results more clearly, let us put some numbers
in. Consider the mixed case with $M_L \sim 1/l_L \sim m_{pl} \sim
10^{19}$ GeV, $M_R \sim  10 - 100$ MeV, and $1/l_R \ll 10^{-34}$
eV. Firstly, note that the current Hubble rate, $H_0 \approx 2/r \sim
 10^{-34}$ eV, as desired. Now consider our source to be the Sun, with
mass $m \sim 10^{66}$ eV. Four-dimensional GR is reproduced at
distances between $r_\textrm{Schw}$ and $10^{21}$ cm. Here, the Schwarzschild radius of the
Sun is given by
$r_\textrm{Schw} \sim 1$ km. The large distance limit,  $10^{21}$ cm, extends well beyond the
size of the solar system, so we have no conflict with experiment.

Of course,  the results (\ref{summary}) were derived by going to
quadratic order in perturbation theory. When $p_* \ll p \ll
p_\textrm{uv}$, we found that the quadratic brane bending term
dominated over the linear terms. We might be worried that this means
we have lost control of perturbation theory. However, we will now
argue that we {\it do} have control of perturbation
theory, because terms beyond quadratic order are always small, right
up to $p_\textrm{uv}$.

Suppose, we have two GN coordinate systems, $x^a$ and $\hat x^a$,
where
\be
\hat x^\mu=x^\mu+\xi^\mu(x, z), \qquad \hat z=z+\xi^z(x, z)
\ee
In the ``unhatted'' coordinates, the brane is positioned at $z=0$. For
simplicity we will demand that $\xi^{\mu}(x, 0)=0$. This just
corresponds to a choice of gauge on the brane. In the ``hatted''
coordinates the brane is positioned at $\hat z=\xi^z(x, 0)=f(x)$.    Non-perturbatively, the
metrics in each coordinate system are related as follows
\be
g_{ab}(x, z)=\hat g_{cd}(\hat x, \hat z) \frac{ \del \hat x^c}{\del  x^{a}}
\frac{ \del \hat x^d}{\del  x^{b}}
\ee
We shall now evaluate the $(\mu \nu)$ component at $z=0$ to derive the
metric on the brane
\be \label{nonpertbrane}
g_{\mu\nu}(x, 0)=\hat g_{\mu\nu}(x, f)+\del_\mu f\del_\nu f
\ee
Suppose we write $\hat g_{\mu\nu}=\bar g_{\mu\nu}+\chi_{\mu\nu}$, and
Taylor expand the right-hand side of (\ref{nonpertbrane}). We get
$g_{\mu\nu}(x, 0)= g_{\mu\nu}^\textrm{quad}(x)+\delta g_{\mu\nu}(x)$, where
\be
g_{\mu\nu}^\textrm{quad}(x)=\bar g_{\mu\nu}(x, 0)+f\bar g_{\mu\nu}'(x, 0)+\half
f^2  \bar g_{\mu\nu} ''(x, 0) +\chi_{\mu\nu}(x, 0)+ f\chi_{\mu\nu}'(x, 0)
+\del_\mu f\del_\nu f 
\ee
and
\be
 \delta g_{\mu\nu}(x) = \sum_{n=3}^{\infty} \frac{f^n}{n!}\bar g_{\mu\nu}^{(n)}(x, 0)+\sum_{n=2}^{\infty}\frac{f^n}{n!}\chi_{\mu\nu}^{(n)}(x, 0)
\ee
Here prime denotes partial differentiation with respect to $z$. $
g_{\mu\nu}^\textrm{quad}(x)$ includes all those terms that were
considered in the analysis to quadratic order. $\delta g_{\mu\nu}(x)$
contains all those terms that were assumed to be of higher order, and
were consequently ignored. We shall now ask whether or not it was safe
to ignore them.

Roughly speaking, we can associate each $z$ derivative with either a
factor of $p$ or a factor of
$1/l$. 	It is obvious that as long as $pf \ll 1$ and $f/l  \ll 1$,
then $\delta g_{\mu\nu}
\ll g_{\mu\nu}^\textrm{quad}$. In the quadratic analysis it is clear
that $pf$ and $f/l$ are indeed small whenever $p \ll p_\textrm{uv}$. We conclude
that it was entirely consistent to ignore all terms beyond quadratic
order.

In analogy with the DGP
model, we shall assume that a break down in perturbation theory gives
rise to strong coupling. To derive the strong coupling scale we
consider a source of mass $m \sim m_{pl}$, and ask at what scale do we
lose perturbative control~\cite{Porrati:massive}. If this scale were $p_*$ we
would be in trouble. We would expect strong coupling below distances of
around $1000$ km. However, we have just shown that we don't expect to lose
control until at least $p_\textrm{uv}$, which typically corresponds to the Planck
scale! These results suggest that there is no strong coupling problem
here, although a more thorough analysis is clearly required.

Finally, note that the mixed case in particular appears to have much in common with the
DGP model~\cite{Dvali:DGP}. The expressions for $r$ and $p_*$
seem to have obvious DGP analogues. The cosmological
behaviour is also very similar (see equation (\ref{appFRW})). One can't
help thinking that this is more than mere coincidence. With this in
mind, consider the brane to be the common boundary to the decaying
bulk and  to the growing bulk. If we were to calculate the boundary
stress-energy tensor on the decaying side, we would expect there to be
divergences. These must be cancelled by boundary counter terms that
will probably take the form of localised curvature on the brane. If
this is indeed the case, it would represent a new mechanism for
obtaining induced curvature on a DGP brane.
\vskip .5in
\centerline{\bf Acknowledgements}
I would like to thank Christos Charmousis and John March-Russell for useful discussions. AP was funded by PPARC.
\medskip

\bibliographystyle{utphys}

\bibliography{irmod}

\end{document}